\documentclass[twocolumn,prb,aps]{revtex4}

\newcommand{\thub}{\tilde t}
\newcommand{\Uhub}{\tilde U}
\newcommand{\Vhub}{\tilde V}

\usepackage{epsfig}

\begin{document}

\title{Dynamical control of correlated states in a square quantum dot}
\author{C.E. Creffield}
\author{G. Platero}

\affiliation{
Instituto de Ciencia de Materiales (CSIC), Cantoblanco, E-28049, Madrid,
Spain}

\date{\today}

\begin{abstract}
In the limit of low particle density, electrons confined to a quantum dot
form strongly correlated states termed Wigner molecules,
in which the Coulomb interaction causes the electrons to become highly
localized in space. By using an effective model of Hubbard-type to
describe these states, we investigate how an oscillatory electric
field can drive the dynamics of a two-electron Wigner molecule held in 
a square quantum dot. We find that, for certain combinations of frequency 
and strength of the applied field, the tunneling between various
charge configurations can be strongly quenched, and we relate
this phenomenon to the presence of anti-crossings in the Floquet quasi-energy
spectrum. We further obtain simple analytic expressions for the location of
these anti-crossings, which allows the effective parameters for a given
quantum dot to be directly measured in experiment, and suggests the exciting
possibility of using ac-fields to control the time evolution of entangled
states in mesoscopic devices.
\end{abstract}

\maketitle

\section{Introduction}
The study of quantum coherent effects in mesoscopic systems, such
as quantum dots (QDs), is
a subject of great current interest, both from the theoretical
point of view, and because of a growing number of possible 
experimental applications.
One of the most notable of these is the swiftly developing field of
quantum computation, in which the coherent manipulation
of entangled quantum states is an essential component. Recent 
experimental successes in detecting Rabi oscillations
in QD systems driven by ac-fields \cite{leo}
have spurred interest in the use of intense ac-fields to coherently manipulate
the time development of electronic states \cite{thz_nature}. An exciting 
possibility is to make use the phenomenon of
{\em coherent destruction of tunneling} (CDT) \cite{hanggi},
in which the tunneling dynamics of a quantum system
become suppressed at certain parameters of the field.
Tuning the driving field thus provides a simple mechanism 
to localize or move charge within the QD on a rapid time-scale
by destroying or restoring the tunneling between regions of the device,
so allowing ac-fields to be used as ``electron tweezers''. 

In this paper we study the use of ac-fields for
this purpose, by investigating the time-dependent behavior
of electrons confined to a two-dimensional QD with a square geometry,
under the influence of a strong driving field. 
We use an effective model of Hubbard-type to describe
the system, which gives a considerable computational
advantage over standard numerical approaches, and also allows us 
to easily include the important effects of the electron correlations 
produced by the Coulomb interaction. By integrating the 
Schr\"odinger equation in time, we find, that for certain values 
of the field, tunneling between different charge configurations 
within the QD can be quenched extremely well. In 
particular we find that tunneling processes parallel to the field can 
be destroyed while transverse tunneling is left unchanged, resulting 
in an effective decoupling between the two halves of the QD.
We explain these findings by making use of the Floquet approach 
\cite{review}, and show that the points at which tunneling is quenched
correspond to anti-crossings between Floquet quasi-energies, 
the locations of which can be found accurately using a perturbational 
method. This significantly clarifies how time-dependent electric fields can 
affect the charge distribution inside a strongly correlated QD, and 
how they can be used for quantum control.

At the low electron densities typically present in QDs,
correlations produced by the Coulomb interaction can significantly
influence the electronic structure. Such strongly correlated problems are
notoriously difficult to treat, and the addition of a time-dependent
field complicates the problem even further. 
When the mean inter-electron separation exceeds a certain critical value, 
however, a considerable simplification occurs, as the Coulomb
interaction dominates the kinetic energy and drives 
a transition to a quasi-crystalline arrangement which 
minimizes the total electrostatic energy. In analogy to the 
phenomenon of Wigner crystallization in bulk two-dimensional 
systems, such a state is termed a {\em Wigner molecule}.
As the electrons in the Wigner state are sharply
localized in space, the system can be naturally and
efficiently discretized by placing 
lattice points just at these spatial locations. 
A many-particle basis can then be constructed by taking Slater 
determinants of single-particle states defined on these lattice
sites, from which an effective Hamiltonian of Hubbard-type can be generated
to describe the low-energy dynamics of the system \cite{jhj_wolf}. 
A major advantage of this technique over standard discretization 
\cite{akbar} schemes,
in which a very large number of lattice points is taken to approximate
the continuum limit, is that
the dimension of the effective Hamiltonian is much smaller 
(typically by many orders of magnitude), which permits the 
investigation of systems which would otherwise be prohibitively complex.   
This approach has proven to be extremely successful in treating a variety of
static problems, including one-dimensional 
QDs \cite{jhj_wolf}, two-dimensional QDs with polygonal
boundaries \cite{creff_wig, creff_mag}, 
and electrons confined to quantum rings \cite{kosk_ring}. 
We further develop this method in this work by including a
time-dependent electric field, and study the temporal dynamics of
the system as it is driven out of equilibrium.

\section{Model and Methods}
\begin{figure}
\centerline{\epsfxsize=60mm \epsfbox{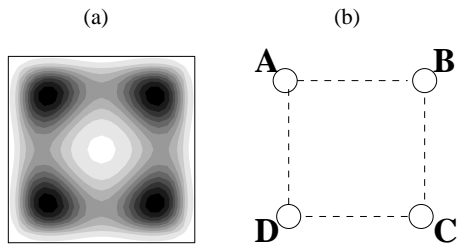}}
\caption{\label{states} (a) Ground-state charge-density for a two-electron square
QD. GaAs material parameters are used, and
the side-length of the QD is 800 nm, placing it in the Wigner regime.
The dark areas indicate peaks in the charge-density.
(b) Lattice points used for the effective lattice-Hamiltonian.}
\end{figure}

We consider a system of two electrons confined to a square QD with a 
hard-wall confining potential -- a simple representation of a two-dimensional 
semiconductor QD. Such a system can be produced
by gating a two-dimensional electron gas confined at a heterojunction
interface, and by placing a gate split into four quadrants over
the heterostructure \cite{4gates}, the potentials at the corners
of the QD can be individually regulated.
In Fig.\ref{states}a we show the ground-state charge-density
obtained from the exact diagonalization of a square QD \cite{creff_wig},
for device parameters placing it
deep in the Wigner molecule regime. It can be seen that the charge-density
is sharply peaked at four points, located close to the vertices of the QD.
This structure arises from the Coulomb interaction between the electrons,
which tends to force them apart into diagonally opposite corners of the dot.
As there are two such diagonal states, degenerate in energy,
we can understand the form of the ground-state by considering it to be 
essentially a superposition of these two states (with a small admixture of
higher energy states). The four points at which the peaks occur
define the sites on which the effective lattice-Hamiltonian operates,
as shown in Fig.\ref{states}b.

We take an effective lattice-Hamiltonian of the form:
\begin{widetext}
\begin{equation}
H =  \sum_{\langle i, j \rangle, \sigma} \left[ \thub \ \left(
c_{i \sigma}^{\dagger} c_{j \sigma}^{ } + H.c. \right) +
\Vhub \ n_i n_j \right]
+ \sum_{i} \left[ \Uhub \ n_{i \uparrow} n_{i \downarrow} +
E_i(t) n_i \right],
\label{hamiltonian}
\end{equation} 
\end{widetext}
where $c_{i \sigma}^{\dagger} / c_{j \sigma}^{ }$ are the
creation/annihilation operators for an electron of spin $\sigma$
on site $i$, $n_{i \sigma} =  c_{i \sigma}^{\dagger} c_{i \sigma}^{ }$,
and $n_i$ is the total charge occupation of site $i$. The quantity
$\thub$ denotes the hopping between adjacent sites, 
and throughout this work we set $\thub$ and $\hbar$ equal to one, and
measure all energies in units of $\thub$.
$\Vhub$ represents the Coulomb repulsion between electrons
occupying neighboring sites, and $\Uhub$ is the
standard Hubbard-$U$ term, giving the energy cost for double-occupation of a 
site. $E_i(t)$ denotes the electric potential at site $i$,
which in general can have a static and a time-dependent component. In experiment, 
static offsets can arise either from deviations of the confining potential
of the QD from the ideal geometry, or by the deliberate application of gating
voltages to the corners of the QD. 
Applying corner potentials in this way could be used to enhance the
stability of the Wigner molecule state, especially if the confining
potential is softer than the hard-wall potential considered here.
Corner gates may also be used to ensure that the multiplet of states included
in this effective lattice-model is well-separated from the other excited
states of the QD system, which will therefore not influence the system's
dynamics. In this work, however, we do not explicitly consider
the effects of static gates, and we neglect the influence of small, accidental
offsets encountered in experiment as we expect them to have only minor effects, 
and indeed may even stabilize CDT \cite{stockberger}. For convenience, we 
consider applying an ac-field aligned with the x-axis of the QD, which can be 
parameterized as:
\begin{equation}
E_A = E_D  = \frac{E}{2} \cos \omega t, \qquad E_B = E_C = - 
\frac{E}{2} \cos \omega t
\end{equation}
where $A,B,C,D$ label the sites as shown in Fig.\ref{states}b.
We emphasize that although we have the specific system of
a semiconductor QD in mind,
the effective-Hamiltonian we are using can describe a wide range of 
physical systems, including $2 \times 2$ arrays of connected 
QDs \cite{kotlyar} and the quantum cellular automata systems 
studied by Lent {\it et al} \cite{lent}, and our results
are thus of general applicability.

\begin{figure} 
\centerline{\epsfxsize=60mm \epsfbox{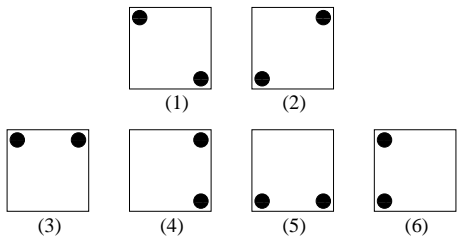}}   
\caption{\label{basis} Schematic representation of the
two-particle basis states for the singlet sub-space of the
Hamiltonian. The ground state of the QD is
approximately a superposition of states (1) and (2).}      
\end{figure}

We study the dynamics of the system by placing it
in a certain initial state, and then integrating this state
in time under the influence of the effective Hamiltonian (\ref{hamiltonian}),
using a fourth-order Runge-Kutta method \cite{creff_ac}.
During the time-evolution, typically of 
the order of $50$ periods of the driving field, we measure physical quantities
such as the particle occupation of the sites $n_i(t)$, and also ensure 
that at all times the unitarity of the wavefunction is accurately preserved. 
As we consider a two-electron system, its eigenstates are
symmetric (antisymmetric) under particle exchange,
corresponding to their singlet (triplet) symmetry.
The Hamiltonian (\ref{hamiltonian}) contains no spin-flip terms
since measurements on semiconductor QDs show that the
spin-flip relaxation time is typically extremely long \cite{fujisawa}, and 
so the singlet and triplet sub-spaces are completely decoupled.
Thus if the initial state possesses a definite parity
this will be retained throughout its time evolution, and
we only need to include states of the same parity in the basis.
We choose to use initial states with singlet symmetry,
which corresponds to the symmetry of the system's ground-state.
Simple state counting reveals that the singlet sub-space 
has a dimension of ten, and can be spanned by the six states
shown schematically in Fig.\ref{basis}, together with the four 
states in which each site is doubly-occupied. 

Since the Hamiltonian (\ref{hamiltonian}) is periodic in time, 
we can use the Floquet theorem to write the 
solutions of the time-dependent Schr\"odinger equation as
$\psi(t) = \exp({-i \epsilon_j t}) \phi_j(t)$, where $\epsilon_j$
is called the quasi-energy, and $\phi_j(t)$ is a function
with the same period as the driving field, called the Floquet
state. This type of expression is familiar in the
context of solid-state physics, where spatial periodicity permits
an analogous rewriting of the spatial wavefunction in terms of
quasi-momenta and Bloch states (Bloch's theorem).
The Floquet states, and their corresponding quasi-energies,
can be obtained from the eigenvalue equation:
\begin{equation}
\left[ H(t) - i \frac{\partial}{\partial t} \right] \phi_j(t) =
\epsilon_j \phi_j(t),
\label{floq_eq}
\end{equation}
which, as we show in the Appendix, is the key to 
obtaining an analytic expression for the quasi-energies by means of
perturbation theory.

Using the Floquet states as a basis, the
time-evolution of a general state driven by the periodic field
may be written as:
\begin{equation}
\psi(t) = \sum_j \ c_j \ \mbox{e}^{-i \epsilon_j t} \phi_j(t),
\label{floq_exp}
\end{equation}
which is formally analogous to the standard expansion
in the eigenvectors of a time-independent Hamiltonian. Indeed, in the
adiabatic limit, $T = 2 \pi / \omega \rightarrow \infty$, the 
quasi-energies evolve to
the eigenenergies, and the Floquet states to the eigenstates.
An important property of expanding in Floquet states (\ref{floq_exp})
is that it provides a extremely valuable separation of time-scales.
Although the Floquet states explicitly depend on time,
they are periodic with the same period as the driving field
and so just influence the dynamics on short time-scales. Consequently, the
long time-scale dynamics of the system is essentially determined by
{\em just} the quasi-energies, and hence evaluating the quasi-energies 
provides a simple and direct way of investigating
the long time-scale behavior of the system.  
In particular, when two quasi-energies are close to degenerate the 
time-scale for tunneling between the states becomes extremely long,
producing the phenomenon of CDT \cite{hanggi, creff_ac}. 

As the ac-field is aligned with the x-axis of the QD, the
Hamiltonian (\ref{hamiltonian}) is invariant under the composite parity
operation $x \rightarrow -x; t \rightarrow t + T/2$. As a result
the Floquet states can also be classified into parity
classes, depending  whether they are odd or even under
this parity operation. Quasi-energies belonging to different
parity classes may cross, but
if they belong to the same class the von Neumann-Wigner
theorem forbids this, and consequently at close
approaches they form anti-crossings instead.
Identifying the presence of crossings and anti-crossings in the 
quasi-energy spectrum provides a necessary (though not sufficient)
condition for CDT to occur.

The locations of these close approaches between quasi-energies may be
found by an analytic approach used by Holthaus \cite{holt_pert}
to treat non-interacting electrons,
and later generalized in Ref.\cite{creff_ac} to include interactions.
In this method the quasi-energies are obtained by first
solving the Floquet equation (\ref{floq_eq}) in the absence of
the tunneling terms, and then performing perturbation
theory with the tunneling terms acting as the ``perturbation''.
A full description of the method is given in the Appendix,
and it has been shown to work extremely well
when the tunneling terms are small in comparison to
the other energy-scales of the problem.

A convenient numerical method to obtain the quasi-energies
and Floquet states is
to diagonalize the unitary time-evolution operator for
one period of the driving field $U(t+T,t)$.
It may be easily shown that the eigenvectors of this operator are
equal to the Floquet states, and its eigenvalues are related to the
quasi-energies via $\lambda_j = \exp[-i \epsilon_j T]$. This method
is particularly well-suited to our approach, as $U(T,0)$ can be
obtained by propagating the unit matrix in time over one period
of the field, using the Runge-Kutta method described earlier.

\section{Results}
\subsection{Non-interacting case}
\label{sec_non}
\begin{figure}
\centerline{\epsfxsize=80mm \epsfbox{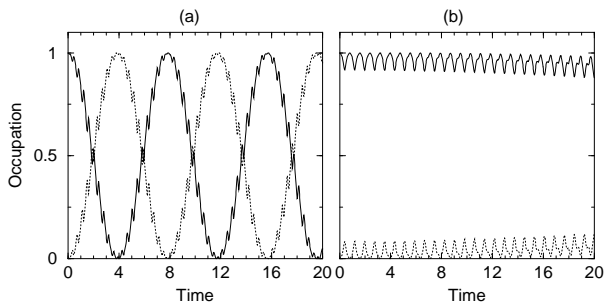}}
\caption{ \label{nonint} Time development of the non-interacting system for $\omega=8$:
(a) electric potential, E = 30.0
(b) E = 19.24.
Solid line indicates the occupation of sites A and D, the dotted line
the occupation of sites B and C.}
\end{figure}

We begin our investigation by first considering the simplest case, 
that of non-interacting electrons ($\Vhub = 0, \ \Uhub = 0$). 
In Fig.\ref{nonint} we show
the time-evolution of the system for two strengths of electric field at
a frequency of $\omega = 8$, in each case using state (6) (see Fig.\ref{basis})
as the initial state. 
A consequence of using this initial state is that
as the ac-field is aligned with the x-axis of the QD,
it does not break the reflection symmetry between the
upper and lower halves of the QD,
and thus throughout the time evolution $n_A = n_D$, and $n_B = n_C$.
For the first case, when the electric potential $E=30.0$, it can be seen that the
occupation of sites cycles between zero and one, as the two electrons
perform spatial Rabi oscillations between the left side of the system
and the right side. The picture in the second case, however, for
$E=19.24$, is
radically different, with the occupation of sites A and D varying 
little from its initial value of one, and sites B and C remaining almost
empty throughout the time evolution. It thus appears that at
this second value of $E$ the tunneling between the
left and right sides has been considerably suppressed.

\begin{figure}
\centerline{\epsfxsize=80mm \epsfbox{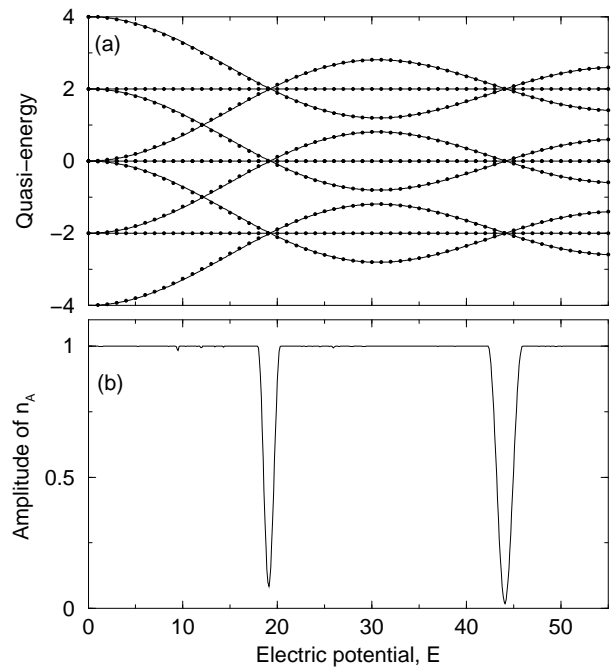}}
\caption{\label{nonint_floq} (a) Quasi-energies of the 
non-interacting system for $\omega=8$:
circles = exact results, lines=perturbative solution.
(b) Amplitude of oscillation of the occupation of site A.}
\end{figure}

To confirm that CDT is occurring, we present in Fig.\ref{nonint_floq} a
comparison of the amplitude of the oscillations of the occupation number of
site A with the quasi-energy spectrum of the system as a function
of the electric potential $E$. It can be clearly seen that for
most field strengths the charge oscillation 
has an amplitude of approximately one, except at sharply defined 
minima where it is heavily quenched. The positions of these minima 
correspond precisely to the locations of exact crossings of the quasi-energies.
Using the perturbative method described in the Appendix
reveals that the quasi-energies fall into
three bands. The central band has quasi-energies of
$\epsilon_{\pm} = \pm 2 J_0(E/\omega)$ (where $J_0$ 
is the Bessel function of the first kind) and $\epsilon_0 = 0$. The upper
band has the same quasi-energies, but increased by a constant amount of $+2$,
and similarly the quasi-energies of the the lower band
are decreased by $2$. Plotting these quantities in
Fig.\ref{nonint_floq}a demonstrates that the agreement of the perturbative result
with the exact results obtained from the diagonalization of $U(T,0)$ is
extremely good, and corroborates our observation of CDT at $E=19.24$,
as at this point
$E / \omega = 2.40$ which is the first zero of $J_0$.  
This dependence of CDT on $J_0$ is familiar from
the behavior of the driven two-level system \cite{hanggi, shirley}, in
which the phenomenon of CDT was first noted.

It is interesting to observe that
the quasi-energy spectrum resembles that of non-interacting electrons
driven by an ac-field in a superlattice \cite{holt_super, holt_pert}, with the 
quasi-energy crossings corresponding to ``miniband collapse''. 
Indeed the lattice model we study can
be considered to be a four-site chain with periodic boundary
conditions in space. The difference between the cases arises, however, 
because in the case of a superlattice
{\em all} intersite tunneling processes are suppressed,
as the electric field is always parallel to the axis of the superlattice.
Although our system is topologically equivalent to a chain,
the fact that it has a two-dimensional geometry means that only
tunneling processes parallel to the field
are suppressed ($A \leftrightarrow B, C \leftrightarrow D$), while 
tunneling in directions perpendicular to the field 
($A \leftrightarrow D, B \leftrightarrow C$) is unaffected.

\subsection{Interacting electrons, no double-occupancy}
\label{sec_Uinf}
We now consider the effect of introducing interactions
between the electrons, 
and begin by taking the Hubbard $U$-term
to be infinitely large -- that is, we work in the
sub-space of states with no double occupation. Our Hilbert
space is thus six-dimensional, and we use as the basis the
states shown in Fig.\ref{basis}.

\begin{figure}
\centerline{\epsfxsize=90mm \epsfbox{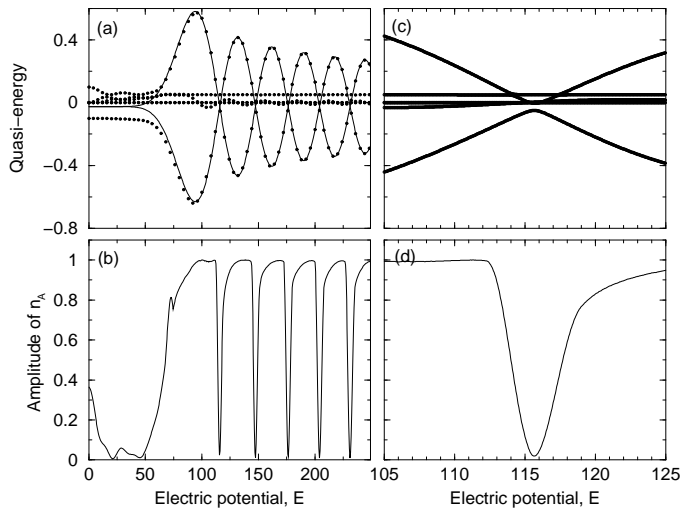}}
\caption{\label{Uinf_floq} (a) Quasi-energies of the system for $\Uhub$ infinite,
$\Vhub=80$ and $\omega=8$:
circles = exact results, lines=perturbative solution [$\pm 2 J_{10}(E/\omega)$].
(b) Amplitude of oscillation of the occupation of site A.
(c) Detail of quasi-energy spectrum, showing an anti-crossing.
(d) Detail of amplitude of oscillations.}
\end{figure}

We show the quasi-energy spectrum of this system, again for a frequency
of $\omega = 8$ in Fig.\ref{Uinf_floq}a. In contrast to the non-interacting
case, we see that the system presents two different regimes of behavior.
The first of these is the weak-field regime,
$E < \Vhub$, at which the driving field does not dominate the dynamics.
In this regime the quasi-energy spectrum, and correspondingly the amplitude of
oscillations, shows little structure, and it is difficult to obtain
analytical results as the perturbational approach is not valid      
when the tunneling terms are comparable in magnitude to the electric field..

The second regime occurs at strong field strengths,
$E > \Vhub$, for which the quasi-energy
spectrum clearly shows a sequence of close approaches. 
In Fig.\ref{Uinf_floq}c we show an enlargement of one of these
close approaches which reveals it to be an {\em anti-crossing}.
Employing the perturbation theory demonstrates that the two quasi-energies 
involved in these anti-crossings are described by $\pm 2 J_n(E/\omega)$, where
$n$ is equal to $\Vhub / \omega$. We may thus think of $n$ as signifying
the number of photons the system needs to absorb to overcome
the Coulomb repulsion between the electrons
occupying neighboring sites. To examine whether these anti-crossings
correspond to CDT, we follow the same procedure as before, and study
the amplitude of oscillations in $n_A$ as a function of $E$, when the system is
initialized in state $(6)$. The results in Fig.\ref{Uinf_floq}b 
and Fig.\ref{Uinf_floq}d strongly confirm 
that tunneling is highly suppressed at the anti-crossings,      
and hence that CDT indeed occurs.

\begin{figure}
\centerline{\epsfxsize=80mm \epsfbox{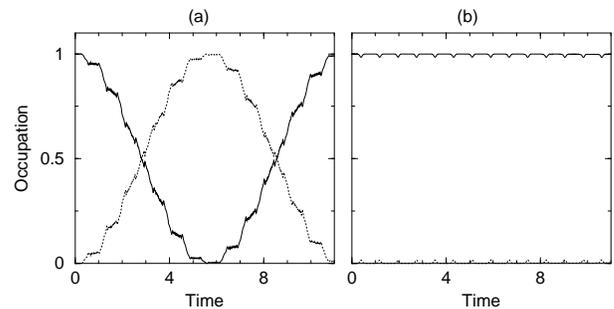}}
\caption{\label{Uinf} Time development of the system for $\Uhub$ infinite, 
$\Vhub=80$ and $\omega=8$:  
(a) electric potential, E = 100.0
(b) E = 115.7.
Solid line indicates the occupation of sites A and D, the dotted line
the occupation of sites B and C.}
\end{figure}

We further show in Fig.\ref{Uinf} the time-dependent number
occupation of the four sites at two values of $E$. In Fig.\ref{Uinf}a
$E$ has a value of 100.0, and it can be clearly seen that the electrons
perform driven spatial Rabi oscillations between the left
side of the QD and the right. Accordingly the occupation of the
sites varies continuously between zero and one. In
Fig.\ref{Uinf}b we show the result of changing the electric
potential to a value of $E=115.7$, which corresponds to the center of
the first anti-crossing. In dramatic contrast to the previous
case, we see that the occupation of site $A$ and $D$ only varies
slightly from unity, while site $B$ and $C$ remain essentially empty
throughout the time-evolution. Only a small amount of charge can transfer 
per period of the driving field
between the left and right sides of the system, producing the small spikes
visible in this figure. The amplitude
of these features is extremely small, however, indicating that
the tunneling between left and right sides has been almost
totally quenched. The efficiency of the quenching depends on
the value of $\Vhub$ which for the case we consider is relatively high.
For smaller values of $\Vhub$, qualitatively the same features occur,
but the efficiency of the quenching is diminished, and the sharpness
of the anti-crossings is reduced.

\subsection{Interacting electrons, double-occupancy permitted}
We now take the most general case, and consider the competition between
the $\Uhub$ and $\Vhub$ terms. Setting $\Uhub$ to a finite value 
means that the four doubly-occupied basis states are no longer
energetically excluded from the dynamics, and accordingly
we must take the full ten-dimensional basis set.

\begin{figure}
\centerline{\epsfxsize=80mm \epsfbox{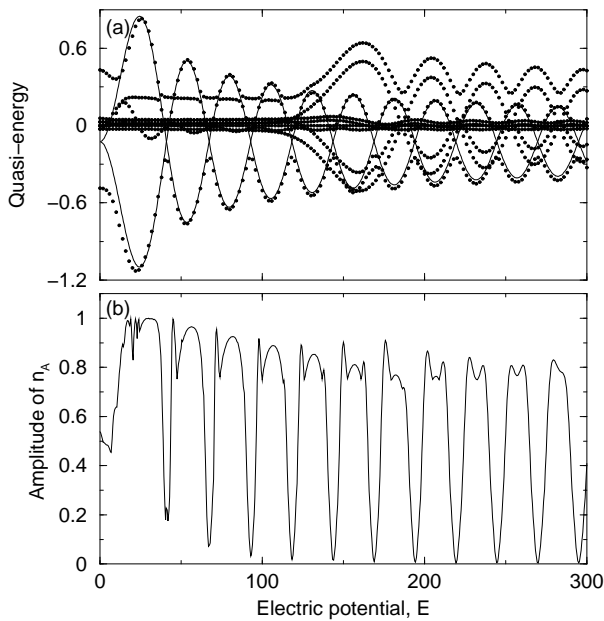}}
\caption{\label{U160_1} (a) Quasi-energies of the system for  $\Uhub=160$ and
$\Vhub=16$, $\omega=8$:
circles = exact results, lines=perturbative solution [$\pm 2 J_{2}(E/\omega)$].
(b) Amplitude of oscillation of the occupation of site A,
with (6) as the initial state.}
\end{figure}

Although it is difficult to obtain precise estimates for the values of
parameters of the effective Hamiltonian, it is clear that
in general $\Uhub > \Vhub$. Accordingly we choose the
parameters $\Uhub = 160, \ \Vhub=16$ to separate
the two energy-scales widely for our investigation.
As before we set the frequency of the ac-field to
$\omega=8$, and in Fig.\ref{U160_1}a we show the quasi-energy
spectrum obtained by sweeping over the field strength. It is immediately
clear from this figure that for electric potentials $E < \Uhub$ the
form of the spectrum is extremely similar to the infinite-$\Uhub$
case. Performing perturbation theory confirms that,
as in the previous case, the behavior of the quasi-energies is
given by $\pm 2 J_n(E/\omega)$ where $n=\Vhub / \omega$.
We show in Fig.\ref{U160_1}b the amplitude
of the oscillations of $n_A$ when the system is
initialized in state (6), which demonstrates that at the locations of the
anti-crossings the tunneling parallel to the field is again quenched.

When the electric potential exceeds the value of $\Uhub$, however,
new structure appears in the quasi-energy spectrum.
A group of four quasi-energies, that for weaker
fields cluster around 
zero, become ``excited'' and make a sequence of anti-crossings
as the field strength is increased. The amplitude of these
oscillations is comparable to the amplitude of the 
two quasi-energies discussed above, but it is clear that
the two sets of anti-crossings are not in phase with
each other. Perturbation theory predicts that these high-field
quasi-energies are given by $\pm 2 J_m(E/\omega)$, where 
$m=(\Uhub-\Vhub)/\omega$, and thus these
anti-crossings arise when the absorption of $m$
photons equates to the electrostatic energy
difference between the two electrons being on neighboring
sites, and doubly-occupying one site. This then indicates
that this structure arises from the coupling of the ac-field 
to the doubly-occupied states.

\begin{figure}
\centerline{\epsfxsize=80mm \epsfbox{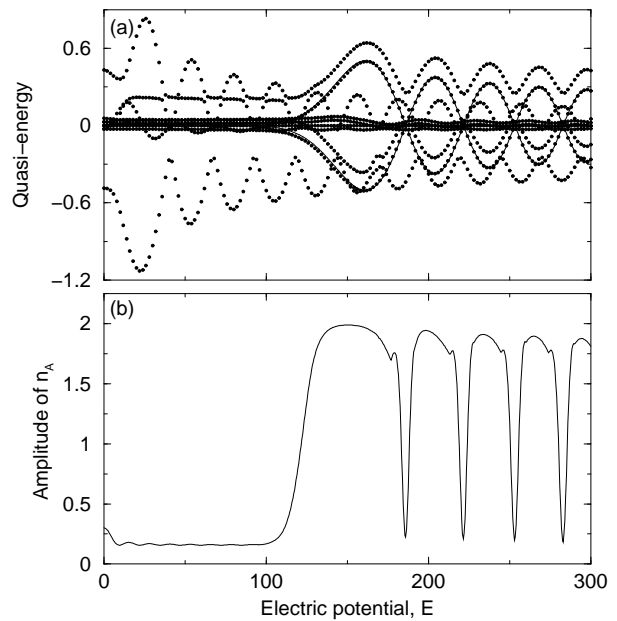}}
\caption{\label{U160_2} (a) Quasi-energies of the system for  $\Uhub=160$ and
$\Vhub=16$, $\omega=8$:
circles = exact results, lines=perturbative solution [$\pm 2 J_{18}(E/\omega)$].
(b) Amplitude of oscillation of the occupation of site A,
with site A doubly-occupied as the initial  state.}
\end{figure}

To probe this phenomenon, we time-evolve the system from
an initial state consisting of {\em two} electrons occupying site $A$. 
In Fig.\ref{U160_2}b it can be seen that for electric potentials
weaker than $\Uhub$ the amplitude of the oscillations
in $n_A$ remains small, and shows little dependence on the
field. As the potential exceeds $\Uhub$, this picture changes, and
the ac-field drives large oscillations in $n_A$, and in fact mainly
forces charge to oscillate between sites $A$ and $B$. At the
high-field anti-crossings, however, the tunneling between
$A$ and $B$ is suppressed, which shuts down this process. Instead,
the only time-evolution that the system can perform consists
of {\em undriven} Rabi oscillations between sites $A$ and $D$,
perpendicular to the field. As these oscillations are undriven they 
have a much longer time-scale than the forced dynamics, and thus during 
the interval over which we evolve the system (approximately 50 periods
of the driving field),
the occupation of $A$ only changes by a small amount, producing the
very sharp minima visible in Fig.\ref{U160_2}b, centered on the roots of 
$J_m(E/\omega)$.

\begin{figure}
\centerline{\epsfxsize=80mm \epsfbox{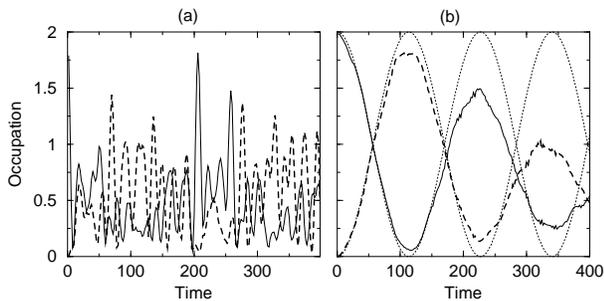}}
\caption{\label{decohere} Time development of the $\Uhub=160$ 
and $\Vhub=16$ system for $\omega=8$:
(a) electric potential, E = 200.0 (off-resonance)
(b) E = 185.8 (on-resonance).
Thick solid line indicates the occupation of site A, the
thick dotted line the occupation of site D.
Dotted lines in (b) show the Rabi oscillations of the
isolated two-site system, Eq.\ref{rabi}.}
\end{figure}

As the tunneling perpendicular to the field is undriven, it is
straightforward to evaluate the time evolution of the initial state,
if we assume that the left side of the QD is
completely decoupled from the right side. The occupation
of sites $A$ and $D$ is then given by:
\begin{equation}
n_A(t) = 1 + \cos \Omega_R t, \quad n_D = 1 - \cos \Omega_R t
\label{rabi}
\end{equation}
where $\Omega_R = 4 \thub^2/(\Uhub - \Vhub)$, which for the parameters
we use, gives a Rabi period of $T_R = 2\pi/\Omega_R=226.19$. In Fig.\ref{decohere}
we display the occupations of sites $A$ and $D$ as a function of time, for 
two values of electric potential. At the first value, $E=200$, tunneling
between the left and right sides of the QD is not quenched, and accordingly
the occupation of the two sites varies rapidly between zero and
two as the electrons are driven by the ac-field around the system.
The second value, $E=185.8$, corresponds to the first
high-field anti-crossing. It can be clearly seen that the charge 
oscillates between sites $A$ and $D$, with a frequency close to $\Omega_R$.
These Rabi oscillations are damped, however, indicating that the isolation
between the left and right sides of the QD is not perfect.
In this sense we can regard the two sites $B$ and $C$ as providing
an environment, causing the quantum system composed of sites
$A$ and $D$ to slowly decohere in time. 
When the tunneling between the left and
right sides of the QD is strong, for example at $E=200$, this
decoherence occurs very rapidly. By moving
to an anti-crossing, however, and suppressing the tunneling,
the rate of mixing between the two sides of the QD can be considerably
reduced, and is just limited by the separation in energy between
the two quasi-energies. Tuning the parameters of the
driving field therefore gives us a simple and
controllable way to investigate how a two-electron
wavefunction can decohere in a QD.

\section{Conclusions}
In this paper we have studied the time-dependent behavior
of a system of two electrons confined to a square QD,
driven by an external ac-field.
By considering the strongly correlated limit of the system,
we are able to use an effective lattice model of just
four sites, which arises from the natural discretization
presented by the system in a Wigner molecule state. We
emphasize, however, that the form of the Hamiltonian we use
is very general, and our results are thus applicable to
a wide variety of mesoscopic systems.
In our effective model, the inter-electron Coulomb interaction
is described by two parameters, $\Uhub$ and $\Vhub$, and the
dynamics of the system consists essentially of tunneling
from corner to corner, along the perimeter of the QD.
We find that when the frequency of the driving field
is in resonance with $\Vhub$ or $(\Uhub - \Vhub)$,
charge moves freely around the system, except at sharply
defined field strengths at which tunneling parallel to
the field is destroyed -- CDT. By using Floquet theory we 
have established that these points correspond to the roots 
of $J_m(E/\omega)$, where $m$ is the order of the resonance
(i.e. $m \omega = \Vhub$ or $m \omega = (\Uhub - \Vhub)$).

When CDT occurs, tunneling parallel to the electric field is suppressed,
and the system is restricted to performing undriven Rabi
oscillations in the perpendicular direction. We can therefore expect 
a radical modification in the frequency and polarization of the EM radiation
emitted by the QD in this situation.
In experiment the phenomenon of CDT should thus
be readily measurable, allowing the measurement of the $\Uhub$ and $\Vhub$
parameters, and hence giving a simple and useful
parameterization of the QD system. The tunability of this
effect, and its ability to discriminate between
states of double-occupation and states
in the single-occupation sub-space, make it an excellent candidate as
a control parameter for manipulating the
dynamics of strongly correlated electrons in mesoscopic systems.

\begin{acknowledgments}
The authors would like to thank John
Jefferson for helpful discussions and comments.
This research was supported by the EU through the TMR programme
``Quantum Transport in the Frequency and Time Domains'', and by the DGES
(Spain) through grant PB96-0875.
\end{acknowledgments}

\appendix
\section{}
The full spin-dependent wavefunction of a two-electron system can 
be factorized into a spinor and a purely spatial wavefunction:
\begin{equation}
\Psi(r_1, \sigma_1; r_2, \sigma_2) = \chi(\sigma_1, \sigma_2) \psi(r_1, r_2),
\end{equation}
where for states of singlet symmetry the spatial wavefunction
is symmetric under particle exchange (and the spinor $\chi$ is
anti-symmetric), while the reverse is true for triplet states.
In this work we consider just the singlet sector of the model, and
the two-particle states shown schematically in
Fig.\ref{basis} represent the spatial components 
of the two-electron wavefunctions. These are formed from
symmetric combinations of single-particle
states defined on the lattice-points $A,B,C,D$, so that,
for example:
\begin{equation}
| 1 \rangle = \frac{\psi_A(r_1) \psi_C(r_2) + \psi_C(r_1) \psi_A(r_2)}
{\sqrt{2}},
\end{equation}
with similar expressions holding for each of the states shown.
An analogous set of states with triplet symmetry can be constructed
by taking anti-symmetric combinations.
To complete the singlet basis we must include the four states in which
a lattice point is doubly occupied, holding one
spin-up electron and one spin-down,
which we label as:
\begin{eqnarray*}
| 7 \rangle =& \psi_A(r_1) \psi_A(r_2) \quad
| 8 \rangle =& \psi_B(r_1) \psi_B(r_2) \cr  
| 9 \rangle =& \psi_C(r_1) \psi_C(r_2) \quad
| 10 \rangle =& \psi_D(r_1) \psi_D(r_2)  
\end{eqnarray*}

Our starting point to obtain approximate expressions for the
quasi-energies is Eq.\ref{floq_eq}. As the Floquet states
are periodic functions of time, it is useful to work in a
Hilbert space of $T$-periodic functions \cite{sambe},
by defining an appropriate scalar product function:
\begin{equation}
\langle \langle \phi_a(t) | \phi_b(t) \rangle \rangle
= \frac{1}{T} \int_0^T \langle \phi_a(t') | \phi_b(t') \rangle dt'
\label{scalar}
\end{equation}
where the single-bracket denotes the usual inner product for the
spatial component of the wavefunctions. The advantage of
working in this extended Hilbert
space is that the Floquet states are {\em stationary} states of
the operator ${\cal H}(t) = H(t) - i (\partial / \partial t)$,
and thus we are able to employ standard stationary perturbation
theory techniques, and avoid the complication of using
time-dependent methods.

We first divide the Hamiltonian (\ref{hamiltonian})
into two parts: $H_t$ which contains all the tunneling
terms, and $H_I$ containing all the interaction terms 
(those involving $\Uhub$, $\Vhub$ and the electric field).
We then find the
eigensystem of the operator ${\cal H}_I(t) = H_I - i (\partial/\partial t)$,
and employ the tunneling Hamiltonian as the
perturbation. By making this division, we can expect
the perturbative result to be good when tunneling is
small in comparison to the other energy scales of the problem,
and conversely, to break down in the limit
of weak fields.

In the basis we have
chosen $H_I$ is diagonal, with entries:
\begin{widetext}  
\begin{equation}
H_I = \mbox{diag}\left( 0, 0, 
\Vhub, \Vhub - E(t), \Vhub, \Vhub + E(t),
\Uhub + E(t), \Uhub - E(t), \Uhub - E(t), 
\Uhub + E(t) \right),
\end{equation}
\end{widetext} 
where $E(t) = E \cos \omega t$,
and thus finding the eigensystem of ${\cal H}_I(t)$ reduces to the straightforward
task of solving the ten first-order differential equations:
\begin{equation}
\left[ (H_I)_{jj} - i \frac{d}{dt} \right] \phi_j(t) =  \epsilon_j \phi_j(t) .
\label{diffeq}
\end{equation}

For $j=1$ and $2$, (\ref{diffeq}) has the trivial
solutions $\phi_j(t) = 1, \ \epsilon_j = 0$. The third and fifth components
also do not have an explicit time dependence, and so have
similarly simple solutions: $\phi_j(t) = \exp[i (\epsilon_j - \Vhub) t]$.
Imposing periodic boundary conditions sets the value of the
quasi-energy, requiring 
$(\epsilon_j - \Vhub) = m \omega$ where $m$ is an integer.
The remaining solutions are all similar in form to each other, and as
an example, $\phi_{10}(t)$ is given by:
\begin{equation}
\phi_{10}(t) = \exp \left[ - i (\Uhub - \epsilon_{10}) t
- i \frac{E}{\omega} \sin \omega t \right] .
\label{eigbess}
\end{equation}
Imposing periodic boundary conditions on the this solution requires
$(\Uhub - \epsilon) = n \omega$, where $n$ is an integer.

The eigenvalues of ${\cal H}_I$ are thus $0$ (with a two-fold
degeneracy), $\Vhub \ \mbox{mod} \ \omega$ (with a four-fold degeneracy),
and  $\Uhub \ \mbox{mod} \ \omega$ (also with a four-fold degeneracy).
These represent the zeroth-order approximations to the quasi-energies
in the perturbational expansion. It can readily be shown that their
associated eigenvectors $\phi_j(t)$ form an orthonormal basis set, 
and, by using standard degenerate perturbation theory, the quasi-energies
can be obtained to first-order by finding the eigenvalues of
the perturbing operator 
$\langle \langle \phi_i | H_t |  \phi_j \rangle \rangle$.
By using the identity:
\begin{equation}
\exp\left[-i \beta \sin \omega t \right] = \sum_{m=-\infty}^{\infty}
J_m (\beta) \exp \left[-i m \omega t \right],
\end{equation}  
to rewrite the eigenvectors which have the form (\ref{eigbess}),
the scalar products (\ref{scalar}) can be evaluated
straightforwardly, allowing the matrix elements of
the perturbing operator to be obtained, and the operator to be
subsequently diagonalized.

For the general case, with $\Vhub = m \omega$
and $\Uhub = n \omega$, the first-order approximation
to the quasi-energies can be readily shown to be $0$ 
(with a four-fold degeneracy), $\pm 2 J_n(E/\omega)$,
and $\pm 2 J_{n-m}(E/\omega)$ (with each state being
two-fold degenerate). The specific cases treated in
Sections \ref{sec_non} and \ref{sec_Uinf} can be treated in
a similar way, giving the perturbative
solutions quoted there.


\begin{thebibliography}{}

\bibitem{leo}
{T.H. Oosterkamp, T. Fujisawa, W.G. van der Wiel, K. Ishibashi,
R.V. Hijman, S. Tarucha and L.P. Kouwenhoven, Nature (London) {\bf 395},
873 (1998).}

\bibitem{thz_nature}
{B.E. Cole, J.B. Williams, B.T. King, M.S. Sherwin and
C.R. Stanley, Nature {\bf 410}, 60 (2001);
D. Vion, A. Aasime, A. Cottet, P. Joyez, H. Pothier, C. Urbina,
D. Esteve and M.H. Devoret, Science {\bf 296}, 886 (2002).}

\bibitem{hanggi}
{F. Grossmann, T. Dittrich, P. Jung and P. H\"anggi,
Phys. Rev. Lett. {\bf 67}, 516 (1991).}

\bibitem{review}
{M. Grifoni and P. H\"anggi, Phys. Rep. {\bf 304}, 229 (1998).}       

\bibitem{jhj_wolf}
{J.H. Jefferson and W. H\"ausler, Phys. Rev. B {\bf 54},
4936 (1996).}  

\bibitem{akbar}
{S. Akbar and I.-H. Lee, Phys. Rev. B {\bf 63}, 165301 (2001).} 

\bibitem{creff_wig}
{C.E. Creffield, W. H\"ausler, J.H. Jefferson and S. Sarkar,
Phys. Rev. B {\bf 59}, 10719 (1999).}

\bibitem{creff_mag}
{C.E. Creffield, J.H. Jefferson, S. Sarkar and D.L.J. Tipton,
Phys. Rev. B {\bf 62}, 7249 (2000).}  

\bibitem{kosk_ring}
{M. Koskinen, M. Manninen, B. Mottelson and S.M. Reimann,
Phys. Rev. B {\bf 63}, 205323 (2001).}

\bibitem{4gates}
{D.G. Austing, T. Honda and S. Tarucha, Semicond. Sci. Technol.
{\bf 12}, 631 (1997).}

\bibitem{stockberger}
{J.T. Stockburger, Phys. Rev. E {\bf 59}, R4709 (1999).}    

\bibitem{kotlyar}
{C.A. Stafford and S. Das Sarma, Phys. Rev. Lett. {\bf 72}, 3590 (1994);
R. Kotlyar and S. Das Sarma, Phys. Rev. B {\bf 55}, R10205,
(1997).}

\bibitem{lent}
{C.S. Lent, Science, {\bf 288}, 159 (2000); G.H. Bernstein,
I. Amlani, A.O. Orlov, C.S. Lent and G.L. Snider,
Nanotechnology {\bf 10}, 166 (1999).}

\bibitem{creff_ac}
{C.E. Creffield and G. Platero, Phys. Rev. B {\bf 65},
113304 (2002).}  

\bibitem{fujisawa}
{T. Fujisawa, D.G. Austing, Y. Tokura, Y. Hirayama and S. Tarucha,
Phys. Rev. Lett. {\bf 88}, 236802 (2002).}

\bibitem{holt_pert}
{M. Holthaus, Z. Phys. B {\bf 89}, 251 (1992).}

\bibitem{shirley}
{J.H. Shirley, Phys. Rev. {\bf 138}, B979 (1965).}

\bibitem{holt_super}
{M. Holthaus, Phys. Rev. Lett. {\bf 69}, 351 (1992).}

\bibitem{sambe}
{H. Sambe, Phys. Rev. A {\bf 7}, 2203 (1973).}

\end{thebibliography}
\end{document}